\documentclass[reprint,superscriptaddress,amsmath, amssymb,aps,prd,notitlepage,longbibliography,floatfix,nofootinbib,onecolumn]{revtex4-1}

\usepackage{amsmath}
\usepackage{lipsum}
\allowdisplaybreaks[4]
\usepackage{cancel}
\usepackage{extarrows}
\usepackage{tensor}     
\usepackage{float}
\usepackage[caption = false]{subfig} 
\usepackage[final]{graphicx}   
\usepackage[
colorlinks=true,        
citecolor=blue,         
linkcolor=blue,         
urlcolor=blue           
]{hyperref}             
\usepackage{bm}         
\usepackage{xcolor}     
\usepackage{lipsum}
\usepackage{color}      
\usepackage[utf8]{inputenc} 
\usepackage[section]{placeins} 
\usepackage{appendix}
\usepackage[capitalise]{cleveref}
\usepackage{units}
\usepackage{dcolumn}
\newcommand{\nc}{\newcommand*}

\nc{\xbar}{\bar{x}}
\nc{\rhoeq}{\rho_{\mathrm{eq}}}
\nc{\zeq}{z_{\mathrm{eq}}}
\nc{\tla}{\tilde{\lambda}}
\nc{\bt}{\beta}
\nc{\dt}{\delta}
\nc{\Dt}{\Delta}
\nc{\vj}{\vec{j}}
\nc{\vl}{\vec{l}}
\nc{\hx}{\hat{x}}
\nc{\hy}{\hat{y}}
\nc{\bj}{\bm{j}}
\nc{\mJ}{\mathcal{J}}
\nc{\mP}{\mathcal{P}}
\nc{\Msun}{M_\odot}
\nc{\app}{\approx}
\nc{\av}[1]{\langle #1 \rangle}
\nc{\eq}[1]{Eq.~\eqref{#1}}
\nc{\al}{\alpha}
\nc{\Xstar}{X_{\ast}}
\nc{\fpbh}{f_{\mathrm{pbh}}}
\nc{\vth}{\vec{\theta}}
\nc{\vla}{\vec{\lambda}}
\nc{\vd}{\vec{d}}
\nc{\Mmin}{M_{\mathrm{min}}}
\nc{\rmd}{\mathrm{d}}
\nc{\mmin}{{m_{\mathrm{min}}}}
\nc{\mmax}{{m_{\mathrm{max}}}}
\nc{\mR}{\mathcal{R}}
\nc{\tmR}{\tilde{\mathcal{R}}}
\nc{\s}{\sigma}
\nc{\ogw}{\Omega_{\mathrm{GW}}}
\nc{\addref}{[\textcolor{red}{add ref}] }
\nc{\Om}{\Omega}
\nc{\gm}{\gamma}
\nc{\Gm}{\Gamma}
\nc{\gpcyr}{\mathrm{Gpc}^{-3}\,\mathrm{yr}^{-1}}
\nc{\Eq}[1]{Eq.~\eqref{#1}}
\nc{\Fig}[1]{Fig.~\ref{#1}}
\nc{\Table}[1]{Table~\ref{#1}}
\nc{\lvc}{LIGO-Virgo-KAGRA collaboration } 
\nc{\Sec}[1]{Sec.~\ref{#1}}
\nc{\eg}{\textit{e.g.~}}
\nc{\SNR}{\mathrm{SNR}}
\nc{\be}{\mathbf{\epsilon}}
\nc{\bn}{\mathbf{n}}
\nc{\bd}{\mathbf{d}}
\nc{\ba}{\mathbf{a}}
\nc{\eps}{\epsilon}
\nc{\bnu}{\mathbf{\nu}}
\nc{\mb}{\mathbf}
\nc{\bbt}{\mathbf{t}}
\nc{\bth}{\mathbf{\theta}}
\nc{\bep}{\mathbf{\epsilon}}
\nc{\uni}{\mathrm{U}}
\nc{\logu}{\operatorname{\mathrm{log-U}}}
\nc{\RN}{\mathrm{RN}}
\nc{\BN}{\mathrm{BN}}
\nc{\GN}{\mathrm{GN}}
\nc{\mcN}{\mathcal{N}}
\nc{\GWB}{\mathrm{GW}}
\nc{\yr}{\mathrm{yr}}
\nc{\Am}{\mathcal{A}}
\nc{\Dm}{\mathcal{D}}
\nc{\Hm}{\mathcal{H}}
\nc{\sovast}{Soviet Ast.}
\newcommand{\dif}{\mathrm{d}}

\nc{\mrm}{\mathrm}
\nc{\BE}{B\scriptsize{AYES}\normalsize{E}\scriptsize{PHEM}\normalsize  }

\nc{\Ostgw}{\Omega_{\mathrm{GW}}^{\mathrm{ST}}}
\nc{\Ottgw}{\Omega_{\mathrm{GW}}^{\mathrm{TT}}}
\nc{\Ovlgw}{\Omega_{\mathrm{GW}}^{\mathrm{VL}}}
\nc{\Oslgw}{\Omega_{\mathrm{GW}}^{\mathrm{SL}}}
\nc{\cosxi}{\beta}

\nc{\gmPL}{\gamma_{\mathrm{PL}}}
\nc{\APL}{A_{\mathrm{PL}}}

\def\({\left(}
\def\){\right)}
\def\[{\left[}
\def\]{\right]}

\def\e{\begin{equation}}
\def\q{\end{equation}}
\def\m{\begin{eqnarray}}
\def\n{\end{eqnarray}}
\nc{\red}[1]{\textcolor{red}{#1}}

\begin{document}

\title{Constraints on the Primordial Black Hole Abundance through Scalar-Induced Gravitational Waves from Advanced LIGO and Virgo's First Three Observing Runs}

\author{Yang Jiang}
\email{jiangyang@itp.ac.cn}
\affiliation{CAS Key Laboratory of Theoretical Physics, 
    Institute of Theoretical Physics, Chinese Academy of Sciences,Beijing 100190, China}
\affiliation{School of Physical Sciences, 
    University of Chinese Academy of Sciences, 
    No. 19A Yuquan Road, Beijing 100049, China}
\author{Chen Yuan}
\email{corresponding author: chenyuan@tecnico.ulisboa.pt}
\affiliation{CENTRA, Departamento de Física, Instituto Superior Técnico – IST, Universidade de Lisboa – UL, Avenida Rovisco Pais 1, 1049–001 Lisboa, Portugal}
\author{Chong-Zhi Li}
\email{corresponding author: lichongzhi@itp.ac.cn}
\affiliation{CAS Key Laboratory of Theoretical Physics, 
    Institute of Theoretical Physics, Chinese Academy of Sciences,Beijing 100190, China}
\affiliation{School of Physical Sciences, 
    University of Chinese Academy of Sciences, 
    No. 19A Yuquan Road, Beijing 100049, China}
\author{Qing-Guo Huang
}
\email{corresponding author: huangqg@itp.ac.cn}
\affiliation{CAS Key Laboratory of Theoretical Physics, 
    Institute of Theoretical Physics, Chinese Academy of Sciences,Beijing 100190, China}
\affiliation{School of Physical Sciences, 
    University of Chinese Academy of Sciences, 
    No. 19A Yuquan Road, Beijing 100049, China}
\affiliation{School of Fundamental Physics and Mathematical Sciences, Hangzhou Institute for Advanced Study, UCAS, Hangzhou 310024, China}


\begin{abstract}
As a promising dark matter candidate, primordial black holes (PBHs) lighter than $\sim10^{-18}M_{\odot}$ are supposed to have evaporated by today through Hawking radiation. This scenario is challenged by the memory burden effect, which suggests that the evaporation of black holes may slow down significantly after they have emitted about half of their initial mass. We explore the astrophysical implications of the memory burden effect on the PBH abundance by today and the possibility for PBHs lighter than $\sim10^{-18}M_{\odot}$ to persist as dark matter. Our analysis utilizes current LIGO-Virgo-KAGRA data to constrain the primordial power spectrum and infer the PBH abundance. We find a null detection of scalar-induced gravitational waves that accompanied the formation of the PBHs. Then we find that PBHs are ruled out within the mass range $\sim[10^{-24},10^{-19}]M_{\odot}$.
Furthermore, we expect that next-generation gravitational wave detectors, such as the Einstein Telescope and the Cosmic Explorer, will provide even more stringent constraints. Our results indicate that future detectors can reach sensitivities that could rule out PBH as dark matter within $\sim[10^{-29}M_{\odot},10^{-16}M_{\odot}]$ in the null detection of scalar-induced gravitational waves. 
\end{abstract}
\maketitle

\section{Introduction} 
The natural of dark matter (DM) is one of the most intriguing puzzles in modern physics. Among all the DM candidates, primordial black holes (PBHs) are not only a promising candidate, but could also serve as the seeds of supermassive black holes, explaining their origin \cite{Bernal:2017nec,Carr:2018rid}. Moreover, PBHs could also explain the gravitational wave (GW) events detected by the \lvc if they constitute a few thousands of the DM \cite{Sasaki:2016jop,Chen:2018czv,Raidal:2018bbj,DeLuca:2020qqa,Hall:2020daa,Bhagwat:2020bzh,Hutsi:2020sol,Wong:2020yig,DeLuca:2021wjr,Bavera:2021wmw,Franciolini:2021tla,Chen:2021nxo,Chen:2024dxh,Andres-Carcasona:2024wqk,Yuan:2024yyo,Huang:2024wse}.

PBHs are formed during the early stages of the Universe, originating from the gravitational collapse of overdense regions that are excessively massive \cite{Zeldovich:1967lct,Hawking:1971ei,Carr:1974nx,Carr:1975qj}. The excessive mass density in these regions is attributed to significant curvature perturbations (see e.g., \cite{Pi:2017gih,Cai:2018tuh,Cotner:2016cvr,Espinosa:2017sgp,Fumagalli:2020adf,Palma:2020ejf,Pi:2021dft,Meng:2022low,Caravano:2024tlp}), which are more pronounced at scales much smaller than those observed in the cosmic microwave background \cite{Planck:2018vyg}. Traditionally, PBHs with masses lighter than $\sim 10^{-18}M_{\odot}$ are thought to have evaporated by today through Hawking radiation, thus limiting their potential as DM candidates \cite{Khlopov:2008qy,Carr:2009jm,Laha:2019ssq,Laha:2020ivk,Saha:2021pqf}. However, recent theoretical advancements have challenged this view. The mechanism responsible for extending the PBH lifetime is known as the memory burden effect \cite{Dvali:2011aa,Dvali:2012en,Dvali:2013eja,Dvali:2015aja,Dvali:2017ktv,Dvali:2017nis,Dvali:2018vvx,Dvali:2018tqi,Dvali:2018xpy,Dvali:2020wft,Dvali:2024hsb}, which significantly suppresses the evaporation process and allow these PBHs to persist as DM candidates up to the present day \cite{Franciolini:2023osw,Thoss:2024hsr,Alexandre:2024nuo,Balaji:2024hpu,Haque:2024eyh}.

Based on a series work by Dvali {\it et al.} \cite{Dvali:2011aa,Dvali:2012en,Dvali:2013eja,Dvali:2015aja,Dvali:2017ktv,Dvali:2017nis,Dvali:2018vvx,Dvali:2018tqi,Dvali:2018xpy,Dvali:2020wft,Dvali:2024hsb}, the memory burden effect proposes a novel perspective on black hole evaporation.  As a black hole evaporates, it accumulates a ``memory'' of the information it has radiated away, effectively slowing its evaporation process. The concept is rooted in the idea that black holes can be viewed as a complex system with a vast number of soft and hard modes, where the soft modes, often referred to as the ``master modes'', control the energy levels of the hard modes, also known as ``memory modes''. The memory modes are responsible for the majority of a black hole's entropy and can encode an exponential amount of information. When a critical number of soft modes are populated, they interact with the hard modes in such a way that the energy level of the latter becomes dependent on the former. This interdependence implies that the memory modes exert a back-reaction on the master mode dynamics, making the evaporation process less straightforward as the black hole loses mass.

Scalar-induced gravitational waves (SIGWs), which originate from the evolution of primordial scalar perturbations in the early Universe are inevitable byproducts of PBH formation \cite{tomita1967non,Matarrese:1992rp,Matarrese:1993zf,Ananda:2006af,Baumann:2007zm,Saito:2008jc,Saito:2009jt}. For review of SIGWs, refer to \cite{Yuan:2021qgz,Domenech:2021ztg}. In \cite{Romero-Rodriguez:2021aws}, the authors search for the SIGWs in the LIGO/VIRGO O1$\sim$O3 data and found no evidence. Hence, they placed constraints on the amplitude of the power spectrum to be $0.02$ for an infinite narrow spectrum at the scale $10^{17}\mathrm{Mpc}^{-1}$, corresponding to PBHs $\sim10^{-20}\Msun$. Later, SIGWs from quasi-extremal PBHs for which the Hawking
evaporation is not efficient are considered in \cite{Franciolini:2023osw}, and the authors studied the expected constraints by the third-generations GW detectors. More recently, the memory burden effect has been considered  for its implications on the role of very light PBHs with $ m \lesssim 10^{-18}\Msun$ as DM. \cite{Thoss:2024hsr,Alexandre:2024nuo} explored the memory effect and argued PBHs lighter than $\sim 10^{-24}\Msun$ could potentially evade evaporation and hence represent all the DM.
Subsequent discussions by \cite{Balaji:2024hpu} proposed that this scenario might result in a Universe dominated by PBHs, where the GWs generated by fluctuations in PBH number density could impose constraints on the suppression of the PBH mass loss rate. And \cite{Haque:2024eyh} studied the dark sector by the decay of PBHs and the allowed parameter space, depending on the strength of the memory burden effect.

In the context of GW astronomy, we focus on SIGWs and we utilize them to constrain the amplitude of the primordial power spectrum.
By analyzing the SIGW signals in the LIGO-Virgo O1$\sim$O3 data, we infer the abundance of PBHs. Anticipating the future, the capabilities of next-generation ground-based GW detectors such as the Cosmic Explorer (CE) and the Einstein Telescope (ET) are particularly exciting. These advanced detectors will offer unprecedented sensitivity, enabling us to probe deeper into the amplitude of the primordial power spectrum and to expect better constraints on the PBH abundance in a wider PBH mass range. Consequently, we estimate the potential constraints on the primordial power spectrum and the corresponding PBH abundance that could be achieved in the absence of detecting SIGWs by CE and ET.

This paper is organized as follows: we begin by reviewing the theoretical framework of SIGWs and the formation of PBHs in Section \ref{section-sigw} and Section \ref{section-pbh} respectively. We then detail our methodology for placing constraints on the primordial power spectrum and the PBH abundance using the data from Advanced LIGO and Advanced Virgo's first three observing runs (LIGO-Virgo O1$\sim$O3) \cite{KAGRA:2021kbb} as well as the expected constraints by future ground-based GW detectors in Section \ref{section-data}. Finally, we conclude with a summary of our results and their implications in Section \ref{section-result}.

\section{Scalar-Induced gravitational waves}\label{section-sigw}
A dimensionless quantity that characterize the isotropic stochastic background is defined as the energy density of GWs per logarithmic frequency interval normalized by the critical energy to close the Universe, such that:
\begin{equation}
    \Omega_\text{GW}(f)=\frac1\rho_c\frac{\dif\rho_\text{GW}}{\dif\ln f},
\end{equation}
with $\rho_c = 3c^2H_0^2/8\pi G$. With regard to our subject, the energy spectrum of SIGW by today is given by \cite{Kohri:2018awv}
\begin{equation}
        	\Omega_{\mathrm{GW}}(k)=\frac{\Omega_r}{6}\left(\frac{g_*}{g_*^0}\right)\left(\frac{g_{*s}}{g_{*s}^0}\right)^{-4/3} \int_{0}^{\infty} \mathrm{d} u \int_{|1-u|}^{1+u} \mathrm{~d} v 	\frac{v^{2}}{u^{2}}\left[1-\left(\frac{1+v^{2}-u^{2}}{2 v}\right)^{2}\right]^{2} \mathcal{P}_{\zeta}(u k) \mathcal{P}_{\zeta}(v k) \overline{I^{2}(u, v)},
        	\label{eq:omega_sigw}
\end{equation}
where the wavelength is related to the frequency by $k=2\pi f$ and $\Omega_r=9\times 10^{-5}$ stands for the energy density of radiation by today. $g_*$ and $g_{*s}$ are the effective degrees of freedom of energy and entropy respectively which are evaluated when the corresponding wavelength reenters the horizon. The upper index $0$ represents the value at present. 

The kernel function, $\overline{I^{2}(u, v)}$, has an analytical expression during the radiation dominated era, which is given by \cite{Kohri:2018awv}:
\begin{equation}
\overline{I^2(u,v)}=\frac{9(u^2+v^2-3)^2}{32u^6v^6}\Bigg\{\Big(-4uv+(u^2+v^2-3) \ln\Big|{3-(u+v)^2\over 3-(u-v)^2}\Big|\Big)^2+\pi^2\left(u^2+v^2-3\right)^2\Theta(u+v-\sqrt{3})\Bigg\}.
\end{equation}

\begin{figure}[ht]
    \centering
    \includegraphics[width=0.5\columnwidth]{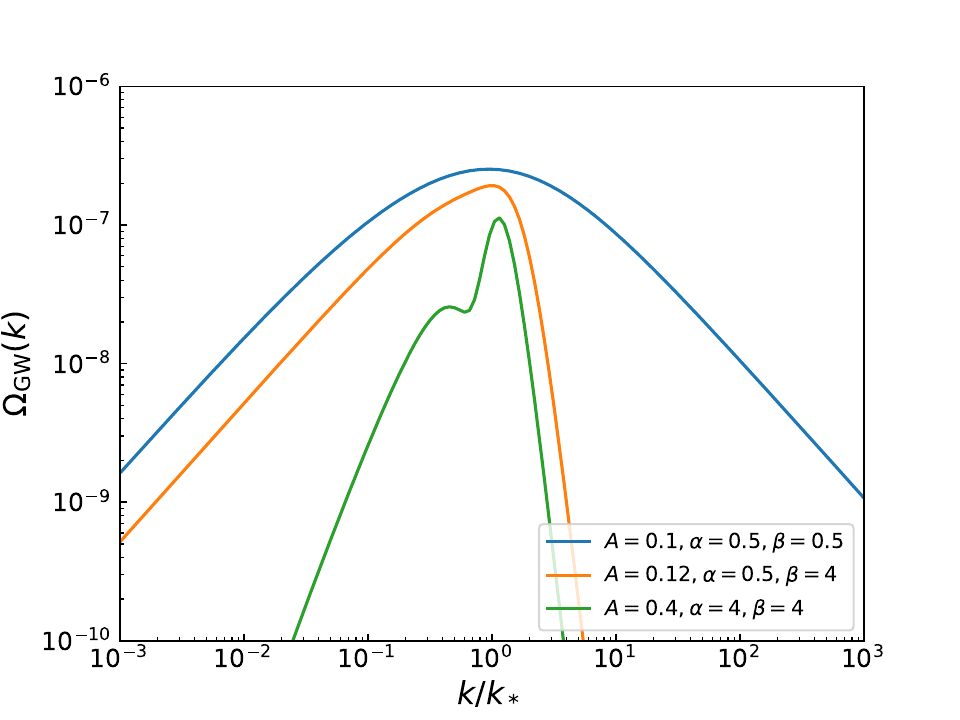}
    \caption{The energy spectrum of SIGWs for specific values of $\alpha$, $\beta$ and $A$.}
    \label{fig:SIGW}
\end{figure}

A broken power-law spectrum that gives a well description of single-field inflation models and curvaton models takes the form
\begin{equation}
    \mathcal{P}_{\zeta}(k) = A \frac{\alpha+\beta} {\beta(k/k_\ast)^{-\alpha}+\alpha(k/k_\ast)^\beta},
\label{BPL}
\end{equation}
where $\alpha>0$ and $\beta>0$ represent the spectral indexes for the growth and decay. The spectrum of SIGWs are demonstrated in Fig.~\ref{fig:SIGW}

\section{formation of primordial black holes}\label{section-pbh}
\begin{figure}[ht]
    \centering
    \includegraphics[width=0.5\columnwidth]{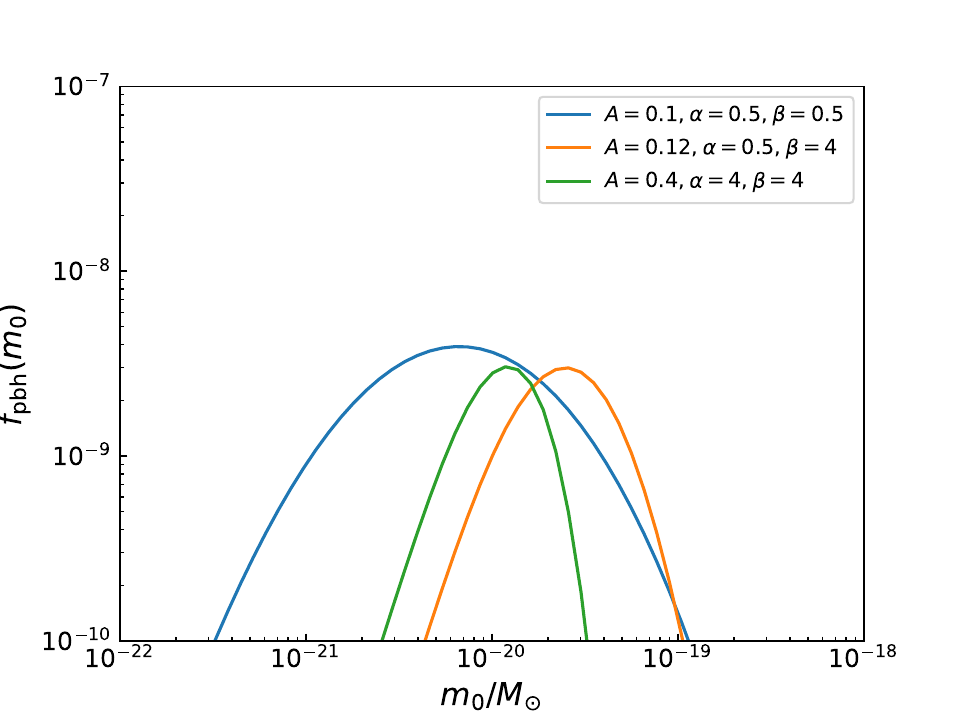}
    \caption{The PBH abundance by today for specific values of $\alpha$, $\beta$ and $A$.}
    \label{fig:pbh}
\end{figure}

In this section, we follow \cite{Ferrante:2022mui,Franciolini:2023pbf,Ianniccari:2024bkh} to calculate the PBH abundance, $\fpbh(m)$, which is normalized as
\begin{equation}
    \int\fpbh(m)\mathrm{d} \log m = \fpbh.
\end{equation}
Here $\fpbh\equiv \Omega_{\mathrm{PBH}} / \Omega_{\mathrm{CDM}}$ denotes the energy fraction of cold dark matter in the form of PBHs and then $\fpbh(m)$ represents the PBH abundance within the mass range $[m,m+\mathrm{d} \log m]$.
The PBH mass fraction is evaluated by integrating the peaked non-linear compact function, which is defined as the mass excess compared to the background value within a given radius \cite{Harada:2015yda}, namely
\begin{equation}\label{beta}
\beta(m)=\int_{\mathcal{C}_c}^{\infty} \mathrm{d} \mathcal{C}\left(r_m\right) \int_{-\infty}^0 \mathrm{~d} \mathcal{C}^{\prime \prime}\left(r_m\right)\frac{m}{M_H} \mathcal{P}\left[\mathcal{C}\left(r_m\right), \mathcal{C}^{\prime \prime}\left(r_m\right)\right],
\end{equation}
where the non-linear compact function during the radiation dominated era is defined as $\mathcal{C}=\mathcal{C}_g-\frac{3}{8}\mathcal{C}_g^2$ with $\mathcal{C}_g=-4/3 r \zeta'(r)$ to be the linear compact function and $\zeta$ is the comoving curvature perturbation. The mass of PBHs is given by \cite{Choptuik:1992jv,Evans:1994pj,Niemeyer:1997mt}
\begin{equation}
    m = M_H \kappa\left(\mathcal{C}(r_m)-\mathcal{C}_c\right)^\gamma,
\end{equation}
where $\kappa=3.3$ and $\gamma=0.36$ \cite{Koike:1995jm} and $M_H\simeq 1.4\times 10^{13}\Msun (k/\mathrm{Mpc}^{-1})^{-2}$ is the horizon mass at PBH formation. The second-order derivative of the compact function in Eq.~(\ref{beta}) is integrated when it is negative since PBHs are formed from local maxima and we have
\begin{equation}
\mathcal{C}^{\prime \prime}\left(r_m\right)=\mathcal{C}_g^{\prime \prime}\left(r_m\right)\left[1-\frac{3}{4} \mathcal{C}_g\left(r_m\right)\right].
\end{equation}
Notice that the region where $\mathcal{C}_g<0$ is the same as $\mathcal{C}<0$ as long as $\mathcal{C}_g<4/3$, known as the type I PBHs. Throughout this paper, we consider type I PBHs.
The threshold, $\mathcal{C}_c$, depends on the shape of the power spectrum \cite{Germani:2018jgr,Musco:2018rwt,Escriva:2019phb}, namely $C_g$ should be a function of $\alpha$ and $\beta$. We adopt the analytical formalism in \cite{Musco:2020jjb} to evaluate $\mathcal{C}_c$.
The probability density function of  $\mathcal{C}_g$ and $\mathcal{C}''_g$ follow joint Gaussian distribution \cite{Ferrante:2022mui,Gow:2022jfb,Ianniccari:2024bkh}:
\begin{equation}\label{2dpdf}
\begin{aligned}
\mathcal{P}\left[-\frac{1}{4} r_m^2 \mathcal{C}_g^{\prime \prime}\left(r_m\right), \mathcal{C}_g\left(r_m\right)\right] & =\frac{\exp \left(-\vec{V}^T \Sigma^{-1} \vec{V} / 2\right)}{2 \pi \sqrt{\operatorname{det} \Sigma}} , \\
\vec{V}^T & =\left[-\frac{1}{4} r_m^2 \mathcal{C}_g^{\prime \prime}\left(r_m\right), \mathcal{C}_g\left(r_m\right)\right], \\
\Sigma & =\left(\begin{array}{cc}
\sigma_2^2 & \sigma_1^2 \\
\sigma_1^2 & \sigma_0^2
\end{array}\right) ,
\end{aligned}
\end{equation}
where the correlations are given by:
\begin{equation}
    \sigma_0^2 = \left\langle \mathcal{C}_g(r_m)^2 \right\rangle ,\quad \sigma_1^2 =-{1\over 4}r_m^2 \left\langle \mathcal{C}''_g(r_m)\mathcal{C}_g(r_m) \right\rangle ,\quad \sigma_2^2 ={1\over 16}r_m^4 \left\langle \mathcal{C}''_g(r_m)^2 \right\rangle.
\end{equation}
These correlations can be computed in the Fourier space. For instance, the linear compact function can be written as  \cite{Ianniccari:2024bkh}
\begin{equation}
    \mathcal{C}_g(\vec{k},r)={4 \over 9}(kr)^2W(k,r)\zeta(\vec{k}),
\end{equation}
where the curvature perturbation is related to the dimensionless power spectrum by $\left\langle \zeta(\vec{k})\zeta(\vec{k'}) \right\rangle=2\pi^2/k^3 \mathcal{P}_{\zeta}(k)\delta(\vec{k}+\vec{k'})$, with $\delta$ to be the Dirac delta function.
We adopt a window function $W(k,r)=\exp(-k^2r^2/4)$ \cite{Ando:2018qdb,Young:2019osy}. The probability density function, $\mathcal{P}\left[\mathcal{C}\left(r_m\right), \mathcal{C}^{\prime \prime}\left(r_m\right)\right]$, can be obtained by changing variables in Eq.~(\ref{2dpdf}).
For PBHs lighter than $\sim 10^{-18}\Msun$, they have evaporated by today through Hawking radiation. However, it is argued that these PBHs might survive from evaporation according to the memory burden effect. The evaporation process is largely suppressed when the PBH has evaporated to about half of its initial mass. Given that the exact value during which the memory burden effect becomes dominated is not known precisely, we estimate the remnant PBH mass by today as $m_0 = q m$. The resulting PBH abundance is then
\begin{equation}
	\fpbh(m_0) =  \frac{q}{\Omega_{\mathrm{CDM}}}\left(\frac{M_{\mathrm{eq}}}{m}\right)^{1 / 2} \beta(m) 
\end{equation}
where the horizon mass at matter-radiation equality is $M_{\mathrm{eq}}\simeq 2.8\times10^{17}\Msun$ and we take $q=1/2$ throughout this paper. We present in Fig.~\ref{fig:pbh} the PBH abundance for specific values of $\alpha$, $\beta$ and $A$.

\section{Hunting for Scalar-induced gravitational waves}\label{section-data}
When GWs propagate through the detector, the perturbations of spacetime will evoke oscillations of two arms in interferometers. This variation of fractional arm length $h(t)=\Delta L(t)/L$ can be read out and it represents the reaction of detector to GWs. Besides, noise component $n(t)$ is inevitable in the facility and it contribute to the strain output of the detectors. It is worth pointing out that due to the weak nature of gravitation, the power of noise is usually much larger than GW signals. Fortunately, however, stochastic GW background (SGWB) will yield non-vanish correlation in the output of detectors. To be specific, an estimator is constructed by the strain output $\tilde{s}_{I,J}(f)$ as follows:
\begin{equation}
    \hat{C}_{IJ}=\frac{2}{T}\frac{\text{Re}[\tilde{s}_I^*(f)\tilde{s}_J(f)]}{\gamma_{IJ}(f)S_0(f)},
\end{equation}
where $S_0(f)=3H_0^2/10\pi^2f^3$ is a normalization factor. $\gamma_{IJ}(f)$ is so-called overlap reduction function between detector pair $I$ and $J$ \cite{PhysRevD.59.102001}. It can be shown that the statistic follows \cite{Romano:2016dpx}
\begin{equation}
    \langle\hat{C}_{IJ}(f)\rangle = \Omega_\text{GW}(f), \qquad
    \sigma_{IJ}^2(f) = \frac{1}{2T\Delta f}\frac{P_I(f)P_J(f)}{\gamma_{IJ}^2(f)S_0^2(f)}.
    \label{eq:estimatorstat}
\end{equation}
$P_{I,J}(f)$ in \Eq{eq:estimatorstat} are power spectral densities of noise in interferometers. $T$ is the obersving time and $\Delta f$ denotes frequency resolution.

\begin{figure}[ht]
    \centering
    \includegraphics[width=0.6\columnwidth]{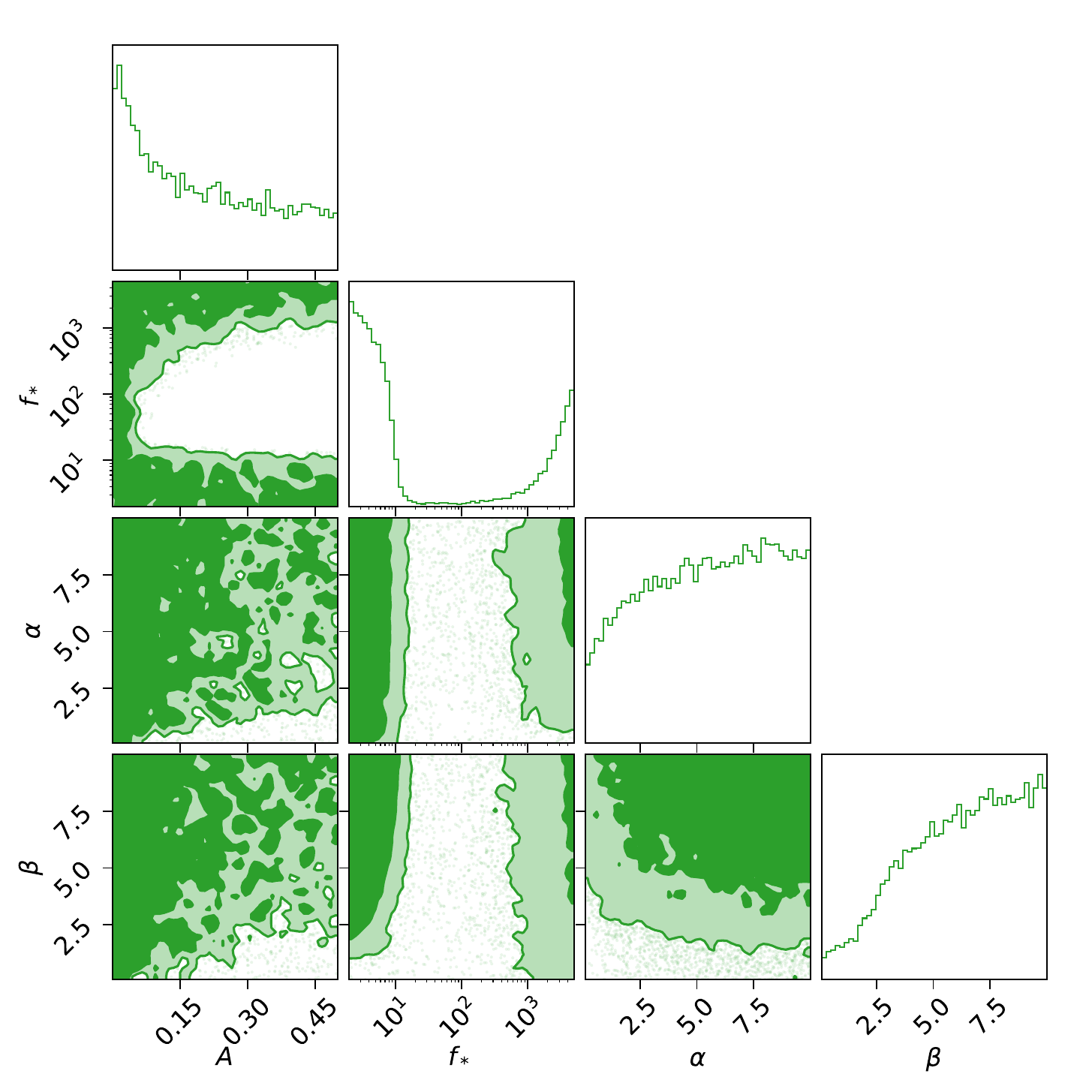}
    \caption{Posterior distributions of the parameters in broken power-law spectrum \Eq{BPL}. Green shaded regions denote $68\%$ and $95\%$ contours.}
    \label{fig:posterior}
\end{figure}

\begin{figure}[htbp!]
\centering
\includegraphics[width = 0.45\textwidth]{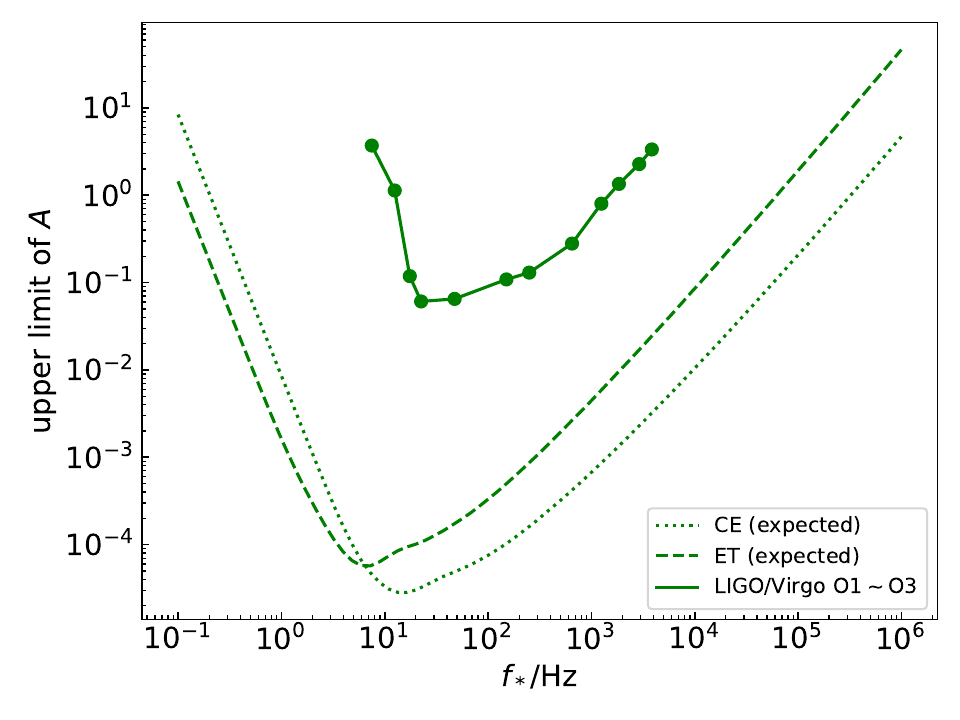}
\includegraphics[width = 0.45\textwidth]{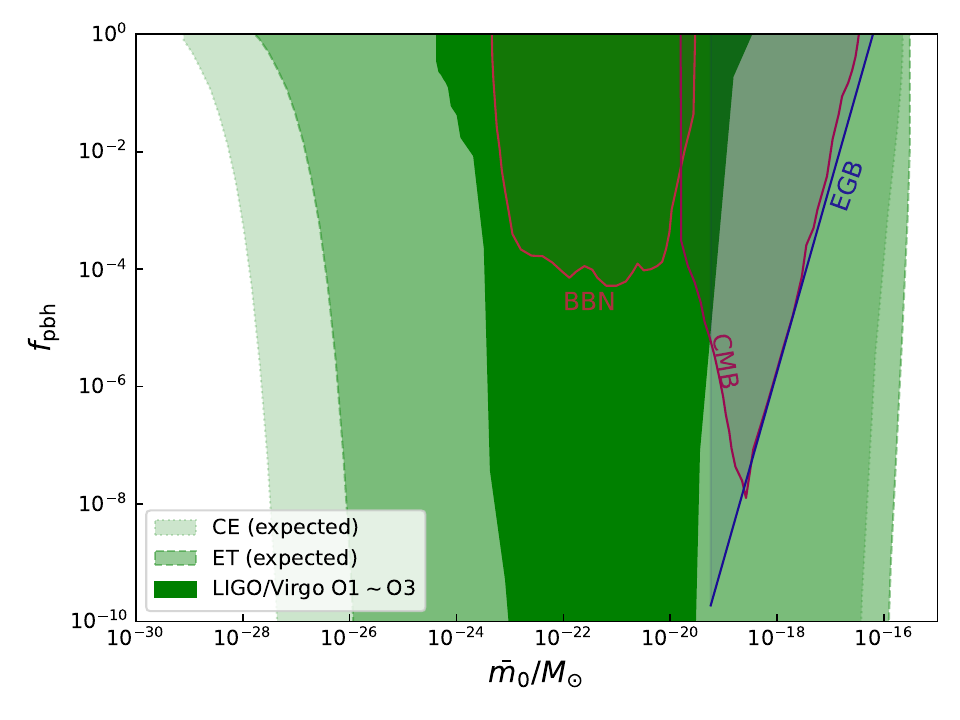}
\caption{Left panel: The constraints on the primordial power spectrum as a function of $f_*$. Right panel: The constraints on the PBH abundance as a function of the averaged PBH mass (our results are labeled green) together with other constraints on $\fpbh$.
\label{fpbhlimit}
}
\end{figure}

We adopt Bayesian approach to implement parameter estimations. After prepossessing, noise can be treated as Gaussian and hence the likelihood is
\begin{equation}
    p(\hat{C}|\bm{\theta})\propto \exp\left[-\frac12\sum_{IJ}\sum_{f}\frac{\left(\hat{C}_{IJ}(f)-\Omega(f;\bm{\theta})\right)^2}{\sigma^2_{IJ}(f)}\right],
\end{equation}
where $\bm{\theta}=\{A,\,f_*,\,\alpha,\,\beta\}$ denotes parameters to be estimated. Bayes factor between pure noise and existence of GW signal
\begin{equation}
    \mathcal{B}_\text{\,noise}^\text{\,model}=\frac{p(\hat{C}\,|\,\text{Model of signal})}{p(\hat{C}\,|\,\text{Pure noise})},
\end{equation}
indicates the possibility of signal existence. Priors in our analysis are listed in \Table{tab:priors}. We set an uniform prior to the amplitude of broken power-law spectrum in \Eq{BPL}. Uniform prior can usually offer a conservative constraint and we set a sufficiently large upper bound so that the corresponding $f_\text{pbh}$ can surpass $O(1)$.
\begin{table}[ht]
    \centering
    \begin{tabular}{p{0.1\columnwidth}<{\centering}p{0.35\columnwidth}<{\centering}p{0.35\columnwidth}<{\centering}}
        \hline\hline
         Parameter & Description & Prior \\
         \hline
         $A$ & Amplitude of broken power-law & $\text{Uniform}[0,\,4]$ \\
         $f_*$ (Hz) & Peak frequency & $\text{LogUniform}[2,\,5000]$ \\
         $\alpha$ & Power growing exponent  & $\text{Uniform}[0.1,\,10]$ \\
         $\beta$ & Power decaying exponent & $\text{Uniform}[0.1,\,10]$ \\
         \hline\hline
    \end{tabular}
    \caption{Prior distributions of the parameters in our estimation.}
    \label{tab:priors}
\end{table}

Besides using the data from LIGO-Virgo O1$\sim$O3, we also make an estimate of the constraints given by the third-generation ground-based GW detectors such as  Einstein Telescope (ET) \cite{Punturo:2010zz,Maggiore:2019uih} and Cosmic Explorer (CE) \cite{LIGOScientific:2016wof}.
Firstly, we evaluate the expected signal-to-noise ratio (SNR) of the SIGWs, $\rho$, as
\begin{equation}
\rho^2 = T \int \mathrm{d} f \frac{S_h(f)^2}{P_n(f)^2 / R(f)^2},
\end{equation}
where $R(f)$ is the detector response \cite{Thrane:2013oya}. For CE and ET, we assume two co-located and co-algined detectors. By setting the threshold SNR for detecting the SIGW to be $\rho_{\mathrm{th}}=1$, we obtain the corresponding results of $A$ as a function of $f_*$.

\section{Results and discussion}\label{section-result}
The result of posteriors is displayed in \Fig{fig:posterior}. We find $\log\mathcal{B}=-1.62$, indicating no evidence for such SGWB. 
And there are no strict restrictions for the parameters $\alpha$ and $\beta$. Once the peak frequency of the spectrum falls in $20\sim200$Hz, the data imposes strong constraints on its amplitude $A$. In this frequency band, the $95\%$ upper limit of the amplitude is $A\sim10^{-1}$, in good agreement with \cite{Romero-Rodriguez:2021aws}.
To evaluate the constraints on the PBH abundance, we compute the averaged PBH mass by today, $\bar{m}_0$ and present a plot of $\fpbh$ as a function of $\bar{m}_0$ in Fig.~\ref{fpbhlimit}. We also include other constraints on $\fpbh$ by extra-galactic Gamma-ray background (EGB) \cite{Carr:2009jm}, the cosmic microwave background (CMB) and the big bang nucleosynthesis (BBN) \cite{Alexandre:2024nuo,Thoss:2024hsr}. The constraints on $\fpbh$ derived from BBN and CMB are rooted in the fact that PBHs in the early universe would have influenced both photons and light elements. However, SIGWs are generated inevitably during the formation of PBHs, representing an independent probes of PBH abundance.


The estimated constraints by CE and ET are demonstrated in Fig.~\ref{fpbhlimit}. For a null detection of SIGW by CE and ET, we evaluate the expected upper limit of $\fpbh$ as a function the averaged PBH mass. The results are shown in Fig.~\ref{fpbhlimit}. It can be seen from Fig.~\ref{fpbhlimit} that the data from LIGO-Virgo O1$\sim$O3 places a constraint on $A$ to $\sim 10^{-1}$ in the most sensitive frequency band (around $30$Hz). We evaluate $\fpbh$ and the averaged PBH mass $\bar{m}_0$ from the posterior and the results show that the data from LIGO-Virgo O1$\sim$O3 can rule out PBHs as DM in the mass range $\sim[10^{-24}, 10^{-19}]\Msun$. Moreover, the constraints by CE and ET are expected to reach $A\sim 10^{-4}$ in the future and can rule out PBH DM in the mass range of $\sim [10^{-29},10^{-16}]\Msun$ in the null detection of SIGWs.

According to \cite{Alexandre:2024nuo}, the mass of PBH evaporating at time $t_H$ can be estimated as
\begin{equation}
    M^{(n)} \sim 10^{-6} \, \text{g} \left( \frac{t_H^{(n)}}{10^{-42} \, \text{s}} \right)^{\frac{1}{3 + 2n}},
\end{equation}
where $n$ is a non-negative integer and $n=0$ corresponds to the standard Hawking radiation case. For the minimal suppression $n=1$, PBHs heavier than $\sim 10^{-27} \Msun$ will survive by today. On the other hand, the lower bound obtained using the LIGO O1-O3 data is $\sim 10^{-24} \Msun$, indicating our results are robust for any positive suppression factor $n$ in the memory burden effects since the lifetime of these PBHs far exceeds the age of the Universe. However, for CE and ET, the lower bounds are $\sim 10^{-29} \Msun$ and $\sim 10^{-28} \Msun$ respectively. Therefore, the constraints near the lower bound by CE and ET are valid for $n \ge 2$.


{\it \textbf{Note added.}} While we were still working on this project, \cite{Kohri:2024qpd} appeared on arXiv. Different from \cite{Kohri:2024qpd}, we use the current available data from LIGO-Virgo O1$\sim$O3 to search for the SIGWs. In addition, \cite{Kohri:2024qpd} and us adopt different models to evaluate the PBH abundance. For example, they use a fixed threshold for an extend PBH mass function, while we consider the change of the threshold with the broadening of the power spectrum (a broad power spectrum leads to more perturbations in the PBH formation and, as a result, a reduction of the threshold). Moreover, the intrinsic non-Gaussianities arising from the non-linear relation between the curvature perturbation and the density contrast (see e.g, \cite{DeLuca:2019qsy,Young:2019yug,Kawasaki:2019mbl}) is considered in this work when computing the PBH abundance.

\section{Acknowledgements.}
This work is supported by the National Key Research and Development Program of China Grant No.2020YFC2201502, grants from NSFC (grant No. 11991052, 12250010, 12475065), Key Research Program of Frontier Sciences, CAS, Grant NO. ZDBS-LY-7009. We acknowledge the use of HPC Cluster of ITP-CAS.
C.Y. acknowledge the financial support provided under the European Union’s H2020 ERC Advanced Grant “Black holes: gravitational engines of discovery” grant agreement no. Gravitas–101052587. Views and opinions expressed are however those of the author only and do not necessarily reflect those of the European Union or the European Research Council. Neither the European Union nor the granting authority can be held responsible for them. We acknowledge support from the Villum Investigator program supported by the VILLUM Foundation (grant no. VIL37766) and the DNRF Chair program (grant no. DNRF162) by the Danish National Research Foundation.

\bibliography{refs}
\end{document}